\documentclass[%
 aip,
 jmp,%
 amsmath,amssymb,
 reprint,%
]{revtex4-1}

\usepackage{graphicx}
\usepackage{dcolumn}
\usepackage{bm}
\usepackage{color, soul}
\begin{document}

\preprint{AIP/123-QED}
\title{Heterogeneous delays making parents synchronized: 
A coupled maps on Cayley tree model}

\author{Aradhana Singh}%
 \email{aradhanas@iiti.ac.in}
\author{Sarika Jalan}%
 \email{sarika@iiti.ac.in}
\affiliation{ 
Complex Systems Lab, Indian Institute of Technology Indore,
IET-DAVV Campus Khandwa Road, Indore 452017, India
}%

\date{\today}

\begin{abstract}
We study the phase synchronized clusters in the diffusively coupled maps on
the Cayley tree networks for heterogeneous delay values. 
Cayley tree networks comprise of two parts: the inner nodes and the boundary nodes. 
We find that heterogeneous delays lead to various cluster states, such as; (a) cluster 
state consisting of inner nodes and boundary nodes, and (b) 
cluster state consisting of only boundary nodes. The former state may comprise of
nodes from all the generations forming self-organized cluster or nodes 
from few generations yielding driven clusters depending upon
on the parity of heterogeneous delay values. 
Furthermore, heterogeneity in delays leads to the lag synchronization between the siblings 
lying on the boundary by destroying the exact synchronization among them. The time lag being equal to the difference in the delay values. 
The Lyapunov function analysis sheds light on the destruction of the exact 
synchrony among the last generation nodes.  
To the end we discuss the 
relevance of our results with respect to their applications in the family business
as well as in understanding the occurrence of genetic diseases.

\end{abstract} 

\maketitle

\begin{quotation}
Many of the real world networks such as river networks, family networks, computer networks and biological networks reflect the tree structure. 
Cayley tree provides a very simple model and thus has been widely studied for instance to model some of the real world networks such as immune network.
Synchronization is an emergent phenomenon where the coupled units adjust their trajectories in some similar manner.  
Our thoughts, action, motion, perceptions all are controlled by the synchronization of neurons in the brain.
Also synchronization plays a very important role in electric power systems, digital telephony, digital audio, video, inscription in telecommunication, flash photography etc and has motivated an intense research on these systems. 
Synchronization can be global as well as local. The local synchronization leads to the cluster formation which is desired in some cases such as in the neural networks and undesired in some cases such as power grid networks, and has thus gained a lot of focus in the last decade.
The finite speed of information transmission leads to time delay, which plays a vital role in synchronization. Moreover in the real world networks  the signal has to travel different distances  and the rate of information transmission can be different for different units, which lead to the heterogeneity in delay values.  
In this paper using delay coupled map model, we study the cluster synchronization on the Cayley tree networks for heterogeneous delay values. 
\end{quotation}

\section{Introduction} 
Cluster synchronization is one of the emergent behaviors observed in the real world networks
\cite{Nature2010,Science2010,SJ_prl2003,Kurths_prl2006}. 
Delay naturally arises in extended systems due to the finite
speed of information transmission \cite{book_delay}. 
A delay gives rise to many new phenomena in dynamical systems such
as oscillation death, enhancement or suppression of synchronization, chimera state, cluster patterns, etc
\cite{osc_death_delay, delay_supress_syn, delay_inhance_syn, Singh2013, delay_coup_osc, chimera, patterns}.
Existance of delay may lead to a comletely different behavior than observed for the undelayed case \cite{book_delay}.
The delay in the model networks can be deliberately implemented in order 
to achieve desired functions such as in case of laser networks for attaining 
secure communication \cite{secure_comm}, whereas in the real world networks the delay
 can be introduced in order to control some of the behaviors such as for the suppression of
undesired synchrony of firing neurons in Parkinson's disease or epilepsy 
\cite{neural_disease_delay1,neural_disease_delay2,neural_disease_delay3}.
The heterogeneous delays present a better model, as communication delays depend on
 the length the signal has to cover and also rate of information transmission from one unit 
to other units \cite{Neural_mul, book_neural1,book_neural2,book_neural3}. Heterogeneity in 
delay adds to the degree of freedom, thus leading to the higher dimensional chaos\cite{book_delay} which 
provides more secured communication\cite{optics}. 

In a recent work, we have demonstrated that heterogeneous delays play a crucial role in formation of
synchronized clusters and mechanism behind the synchronization for coupled maps on 1-d lattice,
small-world, scale-free, random and complete bipartite networks \cite{submitted2013}.
In this paper we study the phase synchronization and lag synchronization on the coupled cayley tree networks with heterogeneous delays.
The Cayley tree is an infinite dimensional regular
graph with an idealized hierarchical structure.
Its idealized hierarchical structure is an ideal model network to investigate
driven patterns in detail. 
Cayley trees have demonstrated their usefulness in
the exact analysis of stability of synchronized 
states \cite{CML_tree}, modeling of immune network with 
antibody dynamics, disease spread \cite{solving_prob, cayley_Ising, cayley_immune} and  
to investigate Bose-Einstein condensation \cite{cayley_bose-condensation}.
Tree structures are found everywhere from the real world networks such as the 
river networks to the technical networks such as power grid networks. 
Tree structure has also been found in the network of sub-fields of physics\cite{sub-field_physics}. 

In the Cayley tree networks of branch ratio $K$ and height $h$
(definition of $h$ excludes the root node), $K^{h}$ nodes lie on the boundary\cite{tree_book}. These nodes are called the leaf or boundary nodes as they do not have children, rest of the nodes are called the inner nodes. There are total
$(K^{h}-1)/(K-1)$ inner nodes in a tree network. Thus in a tree network more than the $50\%$ of the total nodes ($(K^{h+1}-1)/(K-1)$) in the network lie on the boundary.
We demonstrate that inner nodes of the Cayley tree networks are able to get synchronized only for the weaker couplings, while boundary nodes can get synchronized for the stronger couplings as well.  We show that different delay values lead to different phase synchronized patterns consisting of the nodes from the consecutive generations or the nodes from all the generations. The earlier work on the Cayley tree unveils that for homogeneous delays the parents are synchronized only when their children are synchronized \cite{EPJST2013}, 
while this paper reveals study reveals that the presence of heterogeneity in delay may lead to the synchronization between the parent nodes even though there is no synchrony in their children nodes.
Furthermore we observe that there is the lag synchronization between the last 
generation nodes originating from the same parent with the time lag being equal to the difference in the delay values for two nodes. 
\section{Model}
In order to study the phase synchronization on the Cayley tree networks we use well 
known coupled maps model\cite{rev_cml}. The dynamical evolution is defined by following equation,
\begin{equation}
x_i^{t+1} = (1-\varepsilon) f(x_i^t) + \frac{\varepsilon}{k_i} \sum_{j=1}^N A_{ij} f(x_j^{t - \tau_{ij}}).
\label{cml}
\end{equation}
Here we consider network of $N$ nodes and $N_c$ 
 connections between the nodes. Each node of the network
is assigned a dynamical variable $x_i, i = 1, 2, \hdots, N$. 
$A$ is the adjacency matrix with elements
$A_{ij}$ taking values $1$ and $0$ depending upon whether there is a connection 
between $i$ and $j$ or not. The adjacency matrix $A$ is symmetric i.e.$A_{ij} = A_{ji}$.
$ k_{i}$ = $\sum_{j=1}^{N}A_{ij}$ is the degree
of the $i^{th}$ node and $\varepsilon$ is the overall coupling constant.
The delay $\tau_{ij}$ is the time, it takes for the information to reach from a unit $i$ to its neighbor $j$. In case of homogeneous delay $\tau_{ij} = \tau$.
The function $f(x)$ defines the local nonlinear map, as well as the nature
of coupling between the nodes.

In the present investigation we consider networks with 
two delay values $\tau_1$ and $\tau_2$ by randomly making a fraction of connections $f_{\tau_1}$
with delay value $\tau_1$, and another fraction  $f_{\tau_2}$ conducting with the delay $\tau_2$; 
\begin{eqnarray}
f_{\tau_1} = \frac {N_{\tau_1} }{N_c};  \nonumber \\
f_{\tau_2} = \frac {N_{\tau_2} }{N_c}.
\label{frac_delay1}
\end{eqnarray}
where $N_{\tau_1} $ and $N_{\tau_2}$ stands for
the number of connections with delay $\tau_1$ and $\tau_2$ respectively. 
Note that here we are considering heterogeneity in delay ignoring the value of delay, so that when $f_{\tau_1} = f_{\tau_2} = 0.5$ heterogeneity 
in delay is maximum.
Delay in the connections are introduced such that $\tau_{ij}=\tau_{ji}$. 
Depending on the parity of delay values there can be three possibilities \cite{submitted2013}
(a) when both delay values are odd, (b) when one delay value is an odd number and other is an even number, and (c)  when both the delay values are even numbers. 

\begin{figure}
\centerline{\includegraphics[width=8cm, height=4cm]{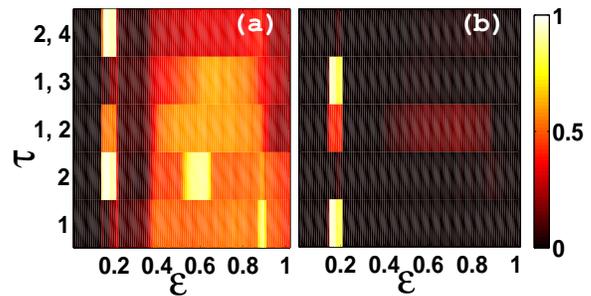}}
\caption{(Color online)Phase diagram demonstrating variation of $f_{inter}$ (a) and $f_{intra}$ (b) in the parameter space of  $\tau_1, \tau_2$ ($\tau$, for homogeneous delays)  and $\epsilon$ for coupled map on 1-d lattice of network size $N=511$ and branch ratio $K=2$. The values on the y axis represent the value of delay.
The grey (color) scale encoding represents the value of $f_{inter}$ and $f_{intra}$. 
The regions, which are black in both the graphs correspond to the state for which there is no cluster formation.
The regions, where both sub-figures have grey (orange) shades, correspond to the state with 
clusters having both inter- and intra-coupling, referred as the mixed cluster state. 
The regions in (a), which are lighter (orange colored) as compared to  
the corresponding $\varepsilon$ and $\tau_{ij}$ ($\tau$) values in (b),  
represents the dominant D phase synchronized clusters state.
The reverse implies to the dominant SO phase synchronized clusters state.
The dark grey (brown)  and dark regions in the (a) and (b) respectively, 
or vice-versa, denote the state where 
a much less clusters are formed.
The figure is obtained by averaging over 20 random initial conditions.}
\label{fig_phase}
\end{figure}
\section{Phase synchronization and synchronized clusters}
Dynamical evolution of the coupled system may lead to the exact synchronization or phase synchronization.
In exact synchronization the dynamical variables for different
 nodes have identical values, whereas in case of phase synchronization 
the dynamical variables for different nodes have some definite relation 
between their phases \cite{book_syn}. 
We consider the phase synchronization  as defined in \cite{SJ_prl2003}. 
Nodes $i$ and $j$ are phase synchronized if their local minima match for all the time
in the interval $T$. 
The pair of nodes which are phase synchronized belong to a single cluster.
Furthermore, lag synchronization represents the state
where one unit lags behind the other unit by a finite time 
$\delta t$, i.e., $x^{t+\delta t}$ = $x_2^{t}$ \cite{book_syn}. 

Depending on the relation between the synchronized clusters and the coupling
between the nodes represented by the adjacency matrix following  phenomena of cluster 
formation have been identified \cite{SJ_prl2003,Singh2013}.

{\it Self-organized clusters}: The nodes of a cluster can be
synchronized because of intra-cluster couplings. 
We refer to this as the self-organized (SO) synchronization and the corresponding 
synchronized clusters as SO clusters. Ideal SO synchronization refer to the state when except those connections which are required to keep the clusters connected, there is no connection out side the clusters. 
Dominant SO synchronization corresponds 
to the state when most of the connections lie inside 
the cluster except a few.

{\it Driven clusters}: The nodes of a cluster can be 
synchronized because of inter-cluster couplings. 
Here the nodes of one cluster are driven by those of
the others. We refer to this as the driven (D) synchronization and
the corresponding clusters as D clusters. 
The ideal D synchronization refers to the state
when clusters do not have any connections within them, and all connections are
outside.
Dominant D synchronization corresponds to 
the state when most of the connections lie outside 
the cluster except a few.

{\it Mixed clusters}: The nodes of a cluster can  
synchronize because of both the inter-cluster couplings and intra-cluster couplings. Such clusters are referred  to as mixed clusters. 

{\it Cluster patterns}: A cluster pattern refers to a particular phase synchronized state, 
which contains information of all the pairs of phase synchronized nodes
distributed in various clusters.
A cluster pattern can be static or dynamic.
Static pattern has all the nodes fixed, except a few floating ones,
{\it in a cluster} with respect to change in time, delay value or initial condition. 
Dynamic pattern changes with time evolution, or 
with initial condition or 
with change in delay value. A change in the pattern refers to the state when members of a cluster get
changed.
Furthermore, patterns can be of D or SO type, which
respectively refers to a particular D or SO phase synchronized state.

The quantities $f_{intra}$ and $f_{inter}$ stand as quantitative measures for 
intra-cluster and inter-cluster couplings; $f_{intra} = N_{intra}/{N_c}$
and $f_{inter} = N_{inter}/{N_c}$,
where $N_{intra}$ and $N_{inter}$ are the numbers 
of intra- and inter- cluster couplings, respectively. In 
$N_{inter}$, coupling between two isolated nodes are not included.

\section{Phase synchronized clusters in the Cayley tree networks:} 
Beginning with random initial conditions, Eq.~\ref{cml} is evolved and the dynamical
behavior of nodes are studied after an initial transient. 
The phase synchronized clusters are considered for $T=100$ 
steps after the initial transient, and  values 
of $f_{inter}$ and $f_{intra}$ are calculated as described earlier. 
We plot the phase diagram demonstrating the variation of $f_{inter}$ and $f_{intra}$ in 
two parameter space of $\varepsilon$ and $\tau$ Fig.~\ref{fig_phase}. 
The phase diagram specifies the parameter space where the synchronized clusters
 are formed and indicates the mechanism behind their formation. In the following we discuss the results for the logistic map ($x(t+1) = \mu x(t)(1-x(t))$) in chaotic regime ($\mu = 4.0$)

The weak coupling range $0.16 \lesssim \varepsilon \lesssim 0.25$ leads to the maximum
 synchronization where almost all the nodes get synchronized.
 In the same coupling range the mechanism of cluster formation changes with change in the
 parity of delay values as observed for the other networks \cite{submitted2013}.
 As the coupling strength increases for the intermediate and strong couplings 
 phase diagram depicts the formation of the ideal D clusters for odd-odd and even-even heterogeneous delays, whereas for the odd-even heterogeneous delays, appearance of the dark grey (red) color at the intermediate couplings exhibit that some of the nodes get synchronized through SO phenomenon. In the same coupling range for homogeneous delay the SO synchronization is not observed.
The study of the clusters based on the structure of this network reveals following behaviors :
\subsection{No synchrony with the parent: D clusters} 
\begin{figure}
\includegraphics[width=6cm,height=3cm]{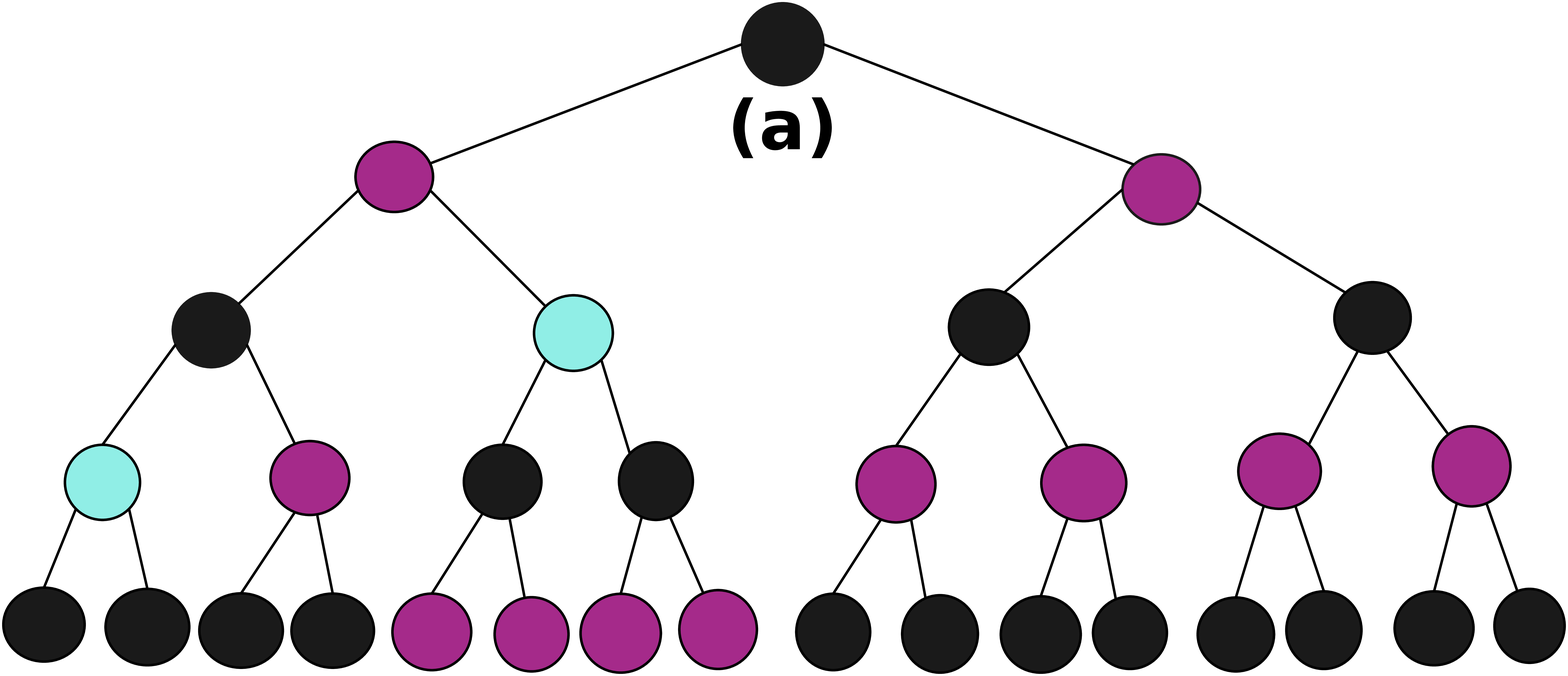}
\includegraphics[width=6cm,height=3cm]{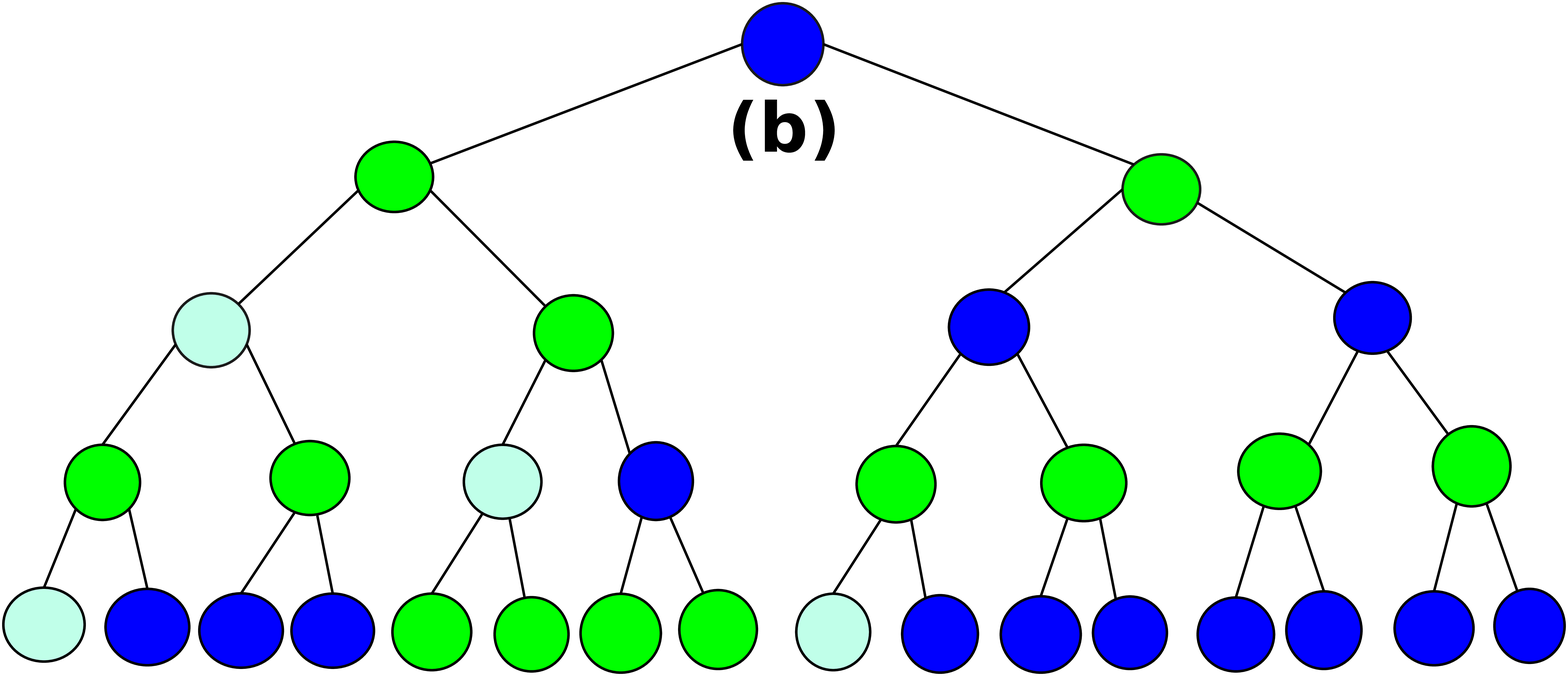}
\caption{(Color online)Phase synchronized clusters with the nodes avoiding immediate
ancestor for Cayley tree networks 
of $N=31$ and $K=2$ (i.e each non-boundary site has 3 neighbors) 
at $\varepsilon=0.18$, for (a) $\tau_1=2$, $\tau_2=4$, and (b) $\tau_1=4, \tau_2=6$.
Closed circle with different  Shades (colors) denote that corresponding nodes belong to same cluster. Open circles represent that the corresponding nodes are not synchronized.
The presented results are true for large network size as well, $N=31$ is considered here for a clear visualization of the phenomena.}
\label{fig_Tree_clus1}
\end{figure}

The phase diagram depicts formation of the ideal D clusters at all the couplings for even-even heterogeneous delays and at the intermediate and strong couplings, for odd-odd heterogeneous delays. Study of the cluster patterns reveals that these clusters correspond to a state
where none of the nodes are synchronized with its immediate ancestor.
Fig.~\ref{fig_Tree_clus1} is pictorial representation of
this state with the ideal D clusters having nodes from the 
alternate generations.
Comparing this with the homogeneous delays case\cite{Singh2013} directs us to the conclude that even delays (either homogeneous or heterogeneous) do not lead to the synchronization between the parent nodes and children nodes for any coupling.
The same  behavior depicted by the even-even heterogeneous and even homogeneous delays can be explained by considering the simple case of weak coupling range, where the even homogeneous delays have been shown to lead the periodic evolution with periodicity depending on the value of delay \cite{Singh2013}.  Since for the 
even-even heterogeneous delay, the difference between the two delays is an 
even number, the introduction of heterogeneity will lead to a similar 
behavior as shown by even homogeneous delays \cite{submitted2013}. 

\subsection{Synchronization of sub-family: SO clusters} 
\begin{figure}[t]
\includegraphics[width=6cm,height=3cm]{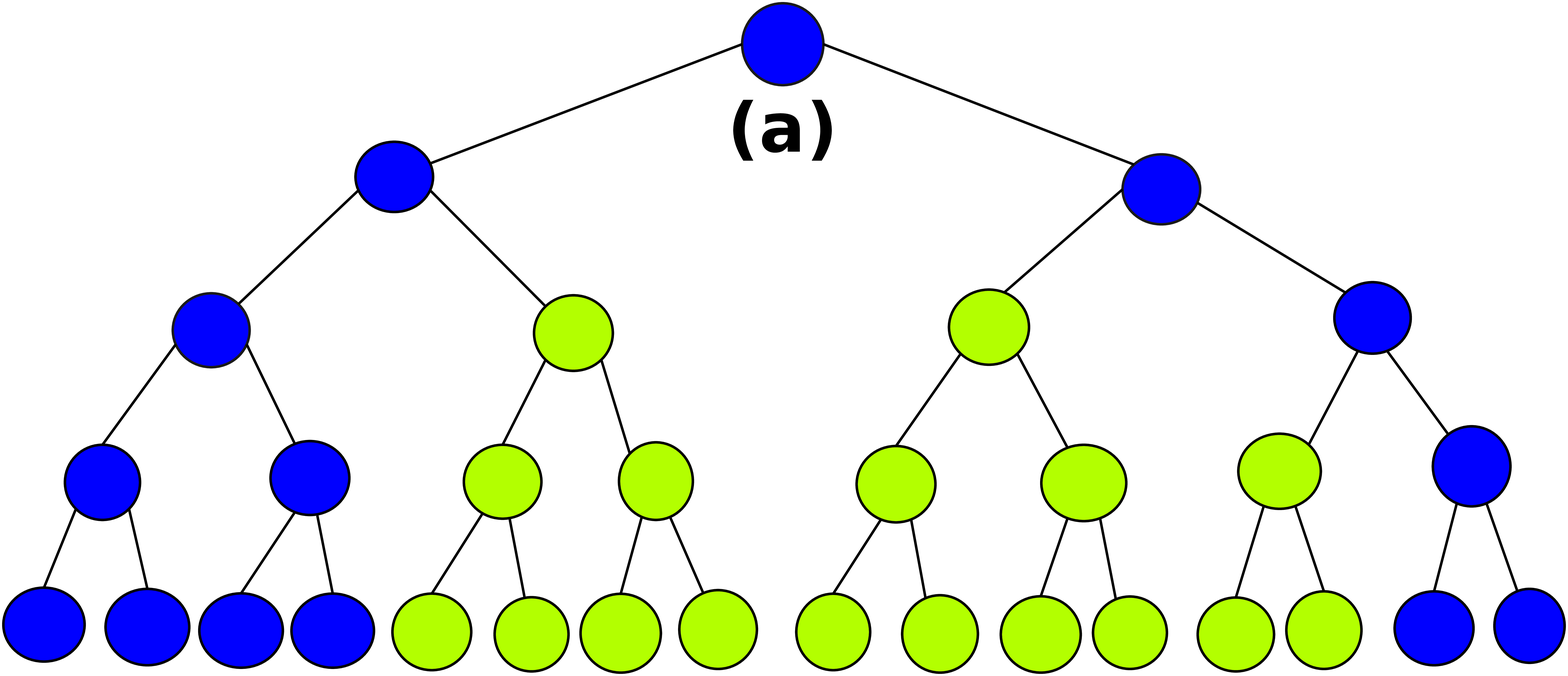}
\includegraphics[width=6cm,height=3cm]{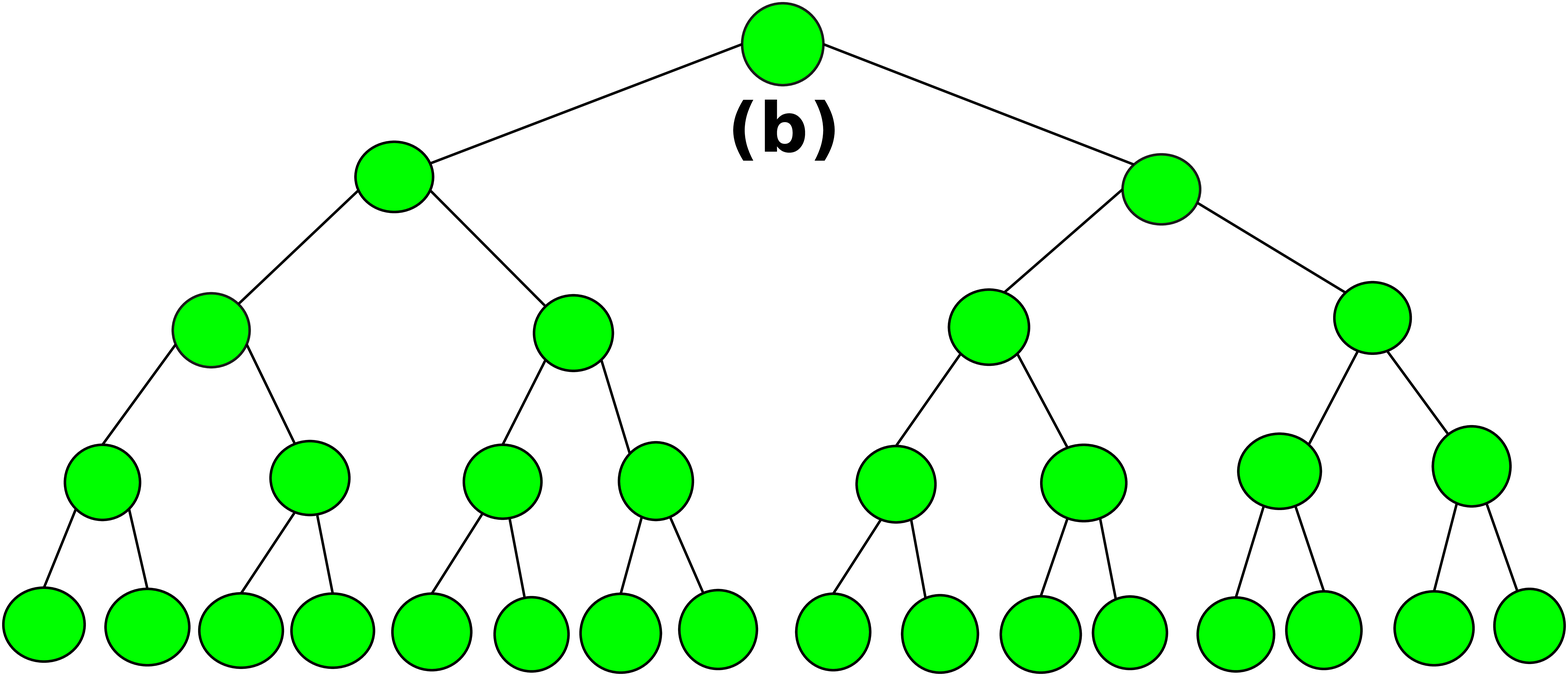}
\caption{(Color online)Phase synchronized clusters consisting of nodes from a sub-family
or nodes from all the generations leading to a global synchronous state 
for Cayley tree networks 
of $N=31$ and $K=2$  
at $\varepsilon=0.18$ for  (a) $\tau_1=1, \tau_2=3$, and (b) $\tau_1=3, \tau_2=5$.
Closed circle with different  Shades (colors) denote that corresponding nodes belong to same cluster. Open circles represent that the corresponding nodes are not synchronized.}
\label{fig_Tree_clus2}
\end{figure}

The odd-odd heterogeneity at the weak coupling range ($0.16 \lesssim \varepsilon \lesssim 0.25$) leads to the formation of two or single SO cluster state (Fig.~\ref{fig_Tree_clus2}). 
These clusters consist of the nodes from all the generation of a family or nodes of a sub-family.  
In the same coupling range the odd homogeneous delays also manifest similar behavior \cite{Singh2013}. Thus the odd delays, divides the network into 
clusters such that  the nodes of all the generations of a family or sub-family get synchronized.
In this coupling range, the odd homogeneous delays are shown to be SO clusters with periodic evolution\cite{Singh2013}, thus the introduction of odd-odd heterogeneity will lead to the similar periodic dynamical evolution being periodic with even periodicity \cite{Singh2013, submitted2013}.

Furthermore a closer look in to the observed clusters reveals that the heterogeneous delay may lead to the following  behaviors, not shown by the homogeneous delays

\subsection{Synchronization of parent nodes}
\begin{figure}
\includegraphics[width=5cm,height=2.5cm]{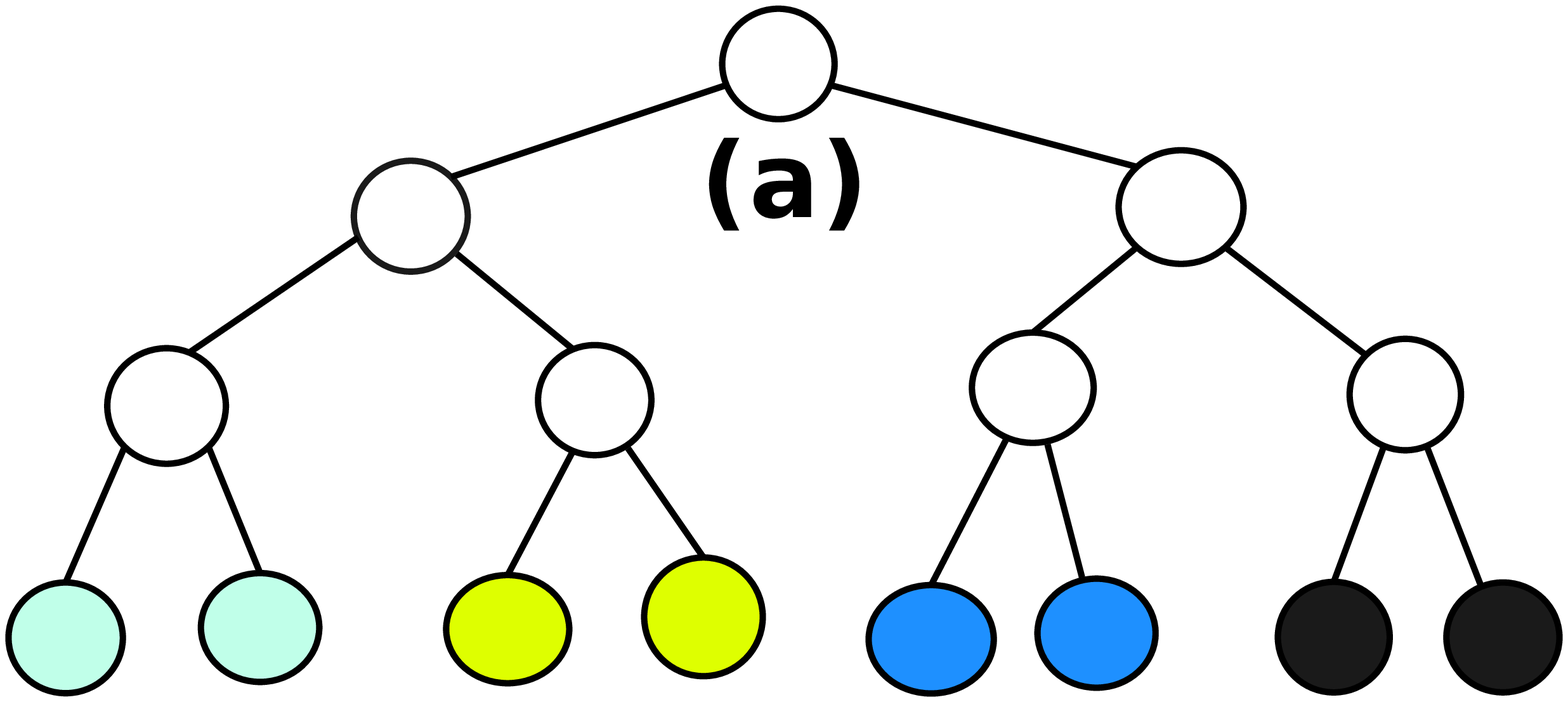}
\includegraphics[width=5cm,height=2.5cm]{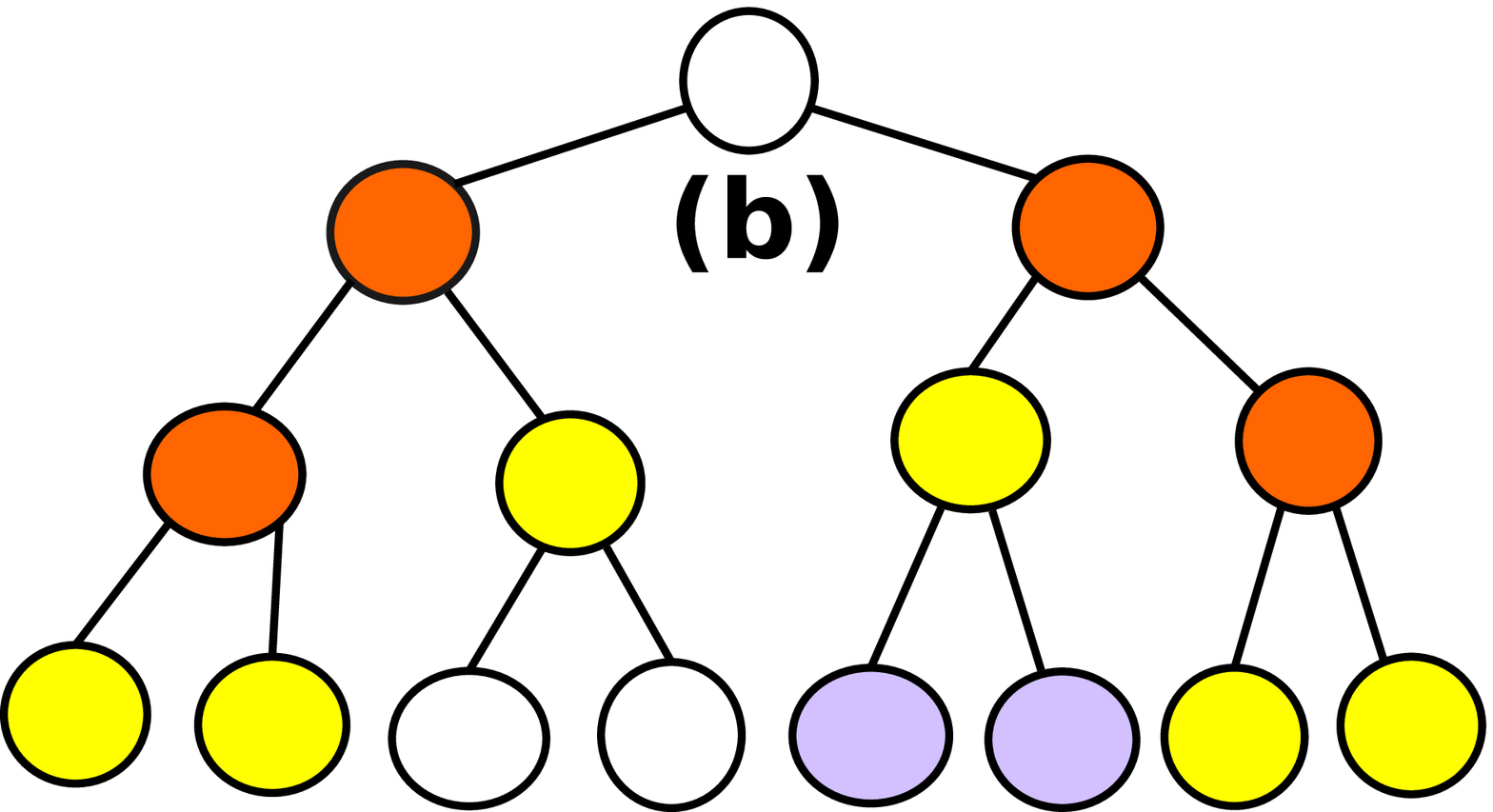}
\caption{(Color online)  illustrating (a) synchronization of the last generation siblings for the homogeneous delay ($\tau =1$) and (b) synchronization of the parent nodes for heterogeneous delays ($\tau_1=1, \tau_2=3$), even though there is no synchronization between their children for Cayley tree networks
of $N=31$, $K=2$ at $\varepsilon=0.7$ . Shades (colors) denote that corresponding nodes belong to same cluster. Open circles represent that the corresponding nodes are not synchronized.}
\label{fig_Tree_clus3}
\end{figure}

\begin{figure} [h]
\centerline{\includegraphics[width=3cm,height=3cm]{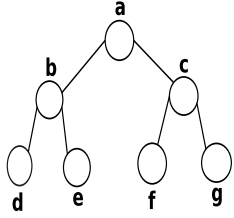}}
\caption{Schematic diagram for the tree network for $K=2$.}
\label{Fig_tree}
\end{figure}

We find that the heterogeneous
delays lead to the synchronization of the parent nodes, even for situations
where their children nodes are not synchronized, a phenomena not observed for the 
homogeneous delay values.
Fig.~\ref{fig_Tree_clus3} plots a demonstration of synchrony in the parent nodes 
accompanied with no synchrony among their children nodes.
In order to understand the origin of  this behavior for heterogeneous delays 
we study the difference variable for two parent nodes, for example nodes b and c in 
Fig.~(\ref{Fig_tree}), given as follows:
\begin{eqnarray}
& & x_b^{t+1} - x_c^{t+1} = (1-\varepsilon)(f(x_b^{t})-f(x_c^{t})) + \nonumber \\
& & \frac{\varepsilon}{K+1} [ {\sum_{p\in S_b}}f(x_{p}^{t-\tau_{bp}})- {\sum_{q\in S_c}}f(x_{q}^{t-\tau_{cq}})+ \nonumber \\
& & (f(x_a^{t-\tau_{ab}}) - f(x_a^{t-\tau_{ac}})) ].
\label{parent_diff}
\end{eqnarray}
where $S_b$ and $S_c$ denote the set of children nodes of $b$ and $c$
respectively.
The coupling terms having the delay
in the right hand side depend on the behavior of 
children nodes as well as of immediate ancestor node of 
$b$ and $c$, respectively. 
Since the immediate ancestor of nodes $b$ and $c$ is common ($a$), for the
homogeneous delay the third 
term in the right hand side cancels out, making 
the synchronization between $b$ and $c$ depend on the synchronization 
between the children nodes only. Thus for the homogeneous delay,
if the children nodes are synchronized then irrespective of the delay value, depending on the coupling strength the parent nodes will also 
get synchronized. 
However, for the heterogeneous delay, the third term in the
right hand side of Eq.~\ref{parent_diff} does not vanish, making the
synchronization of $b$ and $c$
depend on their parent node $a$ as well. Thus for the heterogeneous delay the synchronization between the parent nodes does not solely depend on the synchronization among their children.

\begin{figure}
\includegraphics[width=8cm,height=4.5cm]{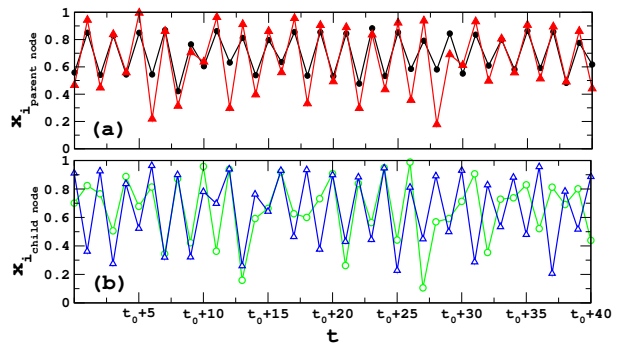}
\caption{(Color online)  illustrating  (a) time evolution of the two parent nodes( closed circle and closed triangle) (b) time evolution of the child node of parent nodes plotted in (a).
 The open circle in (b) correspond to the child node of parent node represented by the closed circle in (a), similarly the open triangle in (b) correspond to the child node of parent node represented by the closed triangle in (a). All the parameters are same as taken in Fig.~\ref{fig_Tree_clus3}}
\label{fig_Tree_clus3_time}
\end{figure}

Furthermore, the D clusters induced by the heterogeneity in delays at 
intermediate couplings are seen to comprise of nodes from the different
generations. Note that for these couplings the D clusters observed for the
homogeneous delay constitute nodes from
the last generation only. The heterogeneity in delays brings nodes from different families together while preserving the underlying mechanism.
Fig.~\ref{fig_Tree_clus3}(b) demonstrates the synchronization of 
different generations for heterogeneous delays. 
 Fig.~\ref{fig_Tree_clus3_time}presents the time evolution of the state of few nodes of Fig.~\ref{fig_Tree_clus3}(b).
This fugure manifests that for the heterogeneous delay, even when the child nodes are not phase synchronized(Fig.~\ref{fig_Tree_clus3_time}(b)) their parent nodes are phase synchronized(Fig.~\ref{fig_Tree_clus3_time}(a)).
In order to find the reason behind the synchronization of inner nodes 
for heterogeneous delay, we study the difference variable for the last 
generation nodes originated from the different parents, for example nodes $d$ and $f$ 
in Fig.~(\ref{Fig_tree}) at $\varepsilon=1$;
\begin{eqnarray}
x_d^{t+1} - x_f^{t+1} &=& (f(x_b^{t-\tau_{bd}})- f(x_c^{t-\tau_{cf}})).
\label{diff_lastgen}
\end{eqnarray}
which in case of homogeneous delays for the chaotic evolution of individual nodes never die for the random initial condition, 
and therefore the synchronization between the last generation nodes from 
different parent nodes is not possible.
As we have already noted that
for homogeneous delay synchronization of the parent nodes depends on the 
synchronization between their children \cite{Singh2013}, thus the parent 
nodes of the last generation nodes (for example $b$ and $c$) can not 
get synchronized for the homogeneous delay, similarly we can explain 
that other ancestors also can not get synchronized.
Thus for homogeneous delay at $\varepsilon=1$ the inner nodes
can not get synchronized, while for the heterogeneous delays as we 
explained above that the behavior of the parent nodes is not completely 
governed by the behavior of the children giving rise
to a possibility for the synchronization of the inner nodes.

To conclude, heterogeneity in delay values makes the synchronization of the parent nodes independent of synchronization among their children nodes and at strong coupling where, homogeneous delay does not lead to the 
synchronization between the inner nodes, heterogeneity in delay 
paves a way to a more coherent behavior.
Although we observe synchronization of the 
inner nodes in the coupling range $0.55\lesssim\varepsilon\lesssim 0.9$,  and 
the analysis carried out here is done for extreme coupling value 
($\varepsilon=1$) which can not be directly applied to other 
$\varepsilon$ values for which terms consisting of local dynamics of 
nodes also appear into the difference variable given by 
Eq.~\ref{diff_lastgen}, but it would have lesser impact on the dynamical 
evolution as compared to the coupled terms for the strong coupling range, and
hence analysis carried out here may stand valid for this range.


\section{Synchronization of boundary nodes} 
As discussed in the introduction section,
in a tree network more than the $50\%$ of the total nodes lie on the boundary, thus in this section we explain the interesting behavior displayed by these
nodes.   
The study of synchronized patterns in presence of heterogeneity in delays reveals 
many different emerging behaviors of these nodes, which are as follows.
  
\subsection{Suppression of the exact synchronization}
At strong couplings, where the heterogeneous delay enhances synchronization in the 1-d lattice, scale-free, random and complete bipartite networks \cite{submitted2013}, for Cayley tree networks there is a suppression in the synchronization. 
Moreover, for the Cayley tree, the homogeneous delays at the strong couplings lead to the exact synchronization among all the nodes originating from the same parent, whereas heterogeneous delay destroys this 
synchrony and distributes them in to different cluster pattern (Fig.~\ref{fig_Tree_clus3}).
 In order to understand D clusters comprising of nodes from different parents in the presence of heterogeneous delay
as compared to the D clusters consisting of nodes from the same parent \cite{EPJST2013} in the presence of the
homogeneous delays, we use Lyapunov function analysis. The Lyapunov function for a pair of 
synchronized nodes can be written as,
\begin{eqnarray}
V_{ij}^{t} &=& [ x_{i}^{t}-x_{j}^{t} ]^{2};
\end{eqnarray}
$V_{ij}^{t}\gtrsim 0$ and the equality holds only when nodes i and j are exactly synchronized.

The above equation in view of Eq.\ref{cml} can be written as follows: 
\begin{eqnarray}
V_{ij}^{t+1} &=& [ (1-\epsilon)( f(x_i^t) - f(x_j^t)) + \nonumber\\
\frac{ \varepsilon}{k_i} \sum_{k=1}^N A_{ik}{g}(x_k^{t - \tau_{ik}})  &- & \frac{\varepsilon}{k_j} \sum_{k=1}^{N}A_{jk} {g}(x_k^{t - \tau_{jk}}) ]^{2}.
\label{lyap_fun}
\end{eqnarray}
For the homogeneous delay the cancellation of the term storing the delay value in Eq.\ref{lyap_fun}  leads to  Synchronization of the last generation nodes belonging to the same parent 

However, for heterogeneity in delay, the siblings connected with edges having different
delay values do not receive same strength of coupling from their parent nodes
and may get distributed into different clusters. 
\begin{figure}
\includegraphics[width=5.7cm,height=2.7cm]{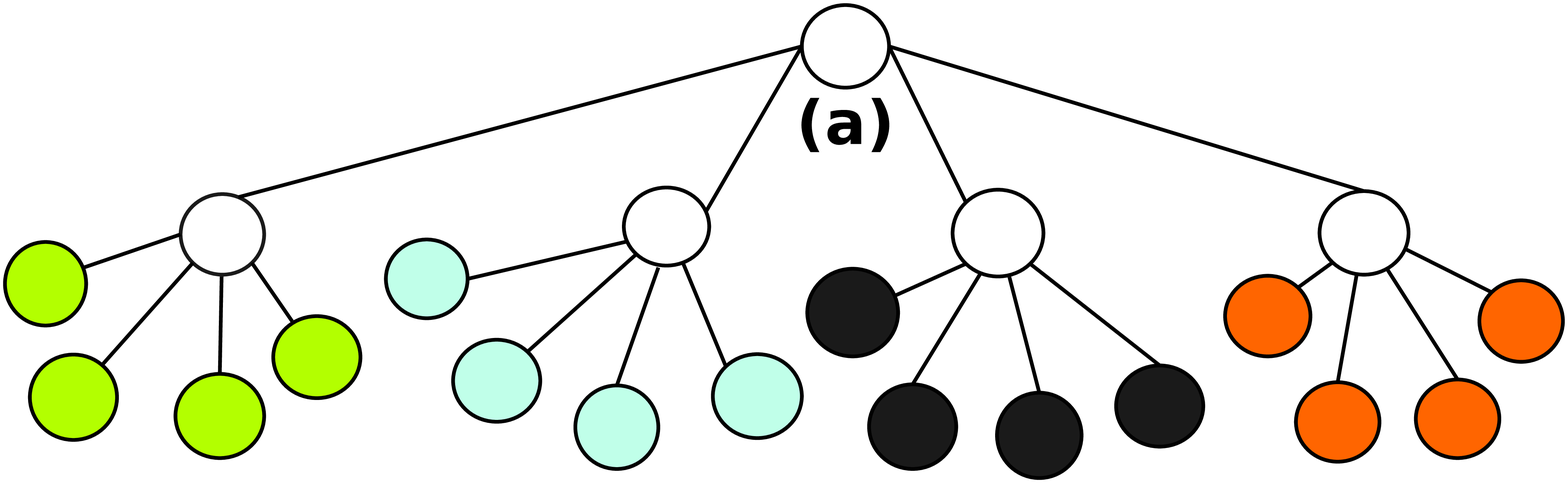}
\includegraphics[width=5.7cm,height=2.7cm]{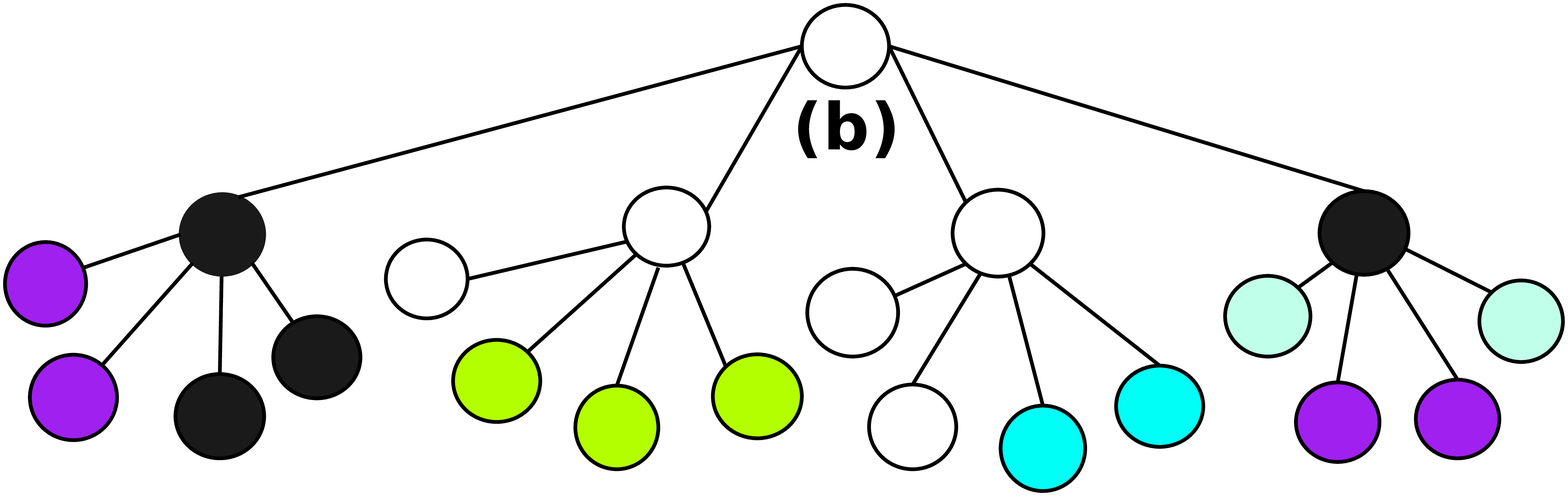}
\includegraphics[width=5.7cm,height=2.7cm]{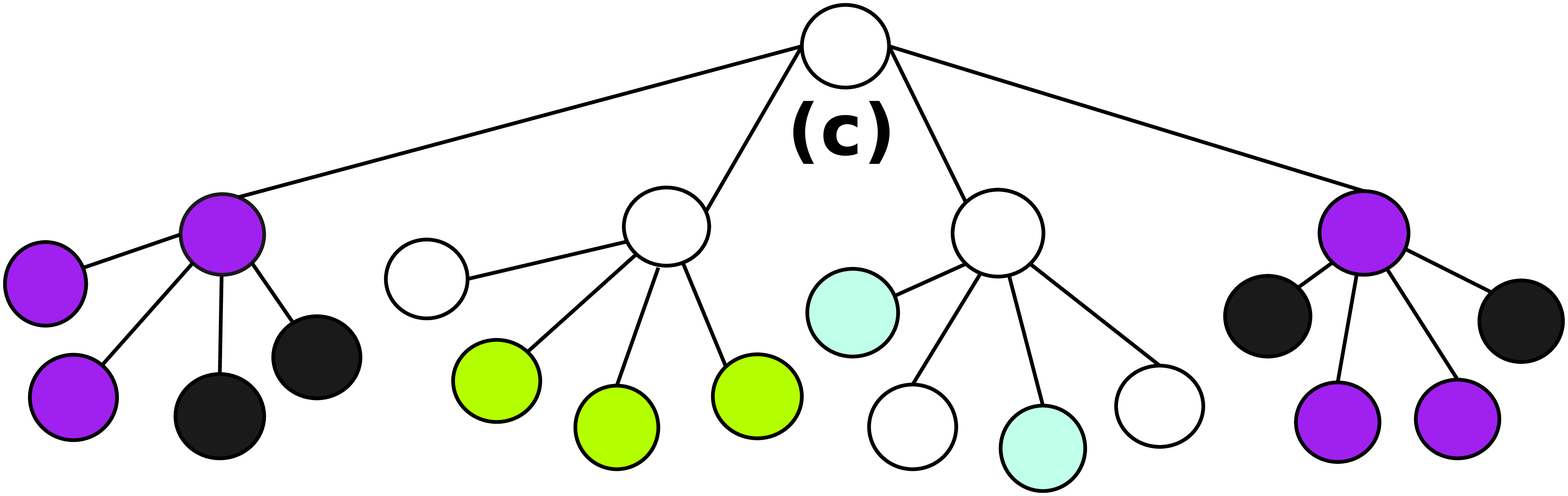}
\caption{(Color online)Phase synchronized clusters (a) comprising last generation siblings for the homogeneous delays ($\tau=1$) and (b) Distribution of the last generation siblings in 
different clusters and synchronization of inner nodes for the heterogeneous delays ($\tau_1=1, \tau_2=2$) for the Cayley tree network of 
$N=21$, $K=4$ at $\varepsilon=0.72$. 
Closed circle with different . Shades (colors) denote that corresponding nodes belong to same cluster. Open circles represent that the corresponding nodes are not synchronized.
The local dynamics of the nodes is governed by the logistic map($x_i^{t+1} = 4 x_i^t(1-x_i^t)$).}
\label{fig_Tree_clus}
\end{figure}

\subsection{Occurrence of lag synchronization}
\begin{figure}[t]
\centerline{\includegraphics[width=8cm,height=6cm]{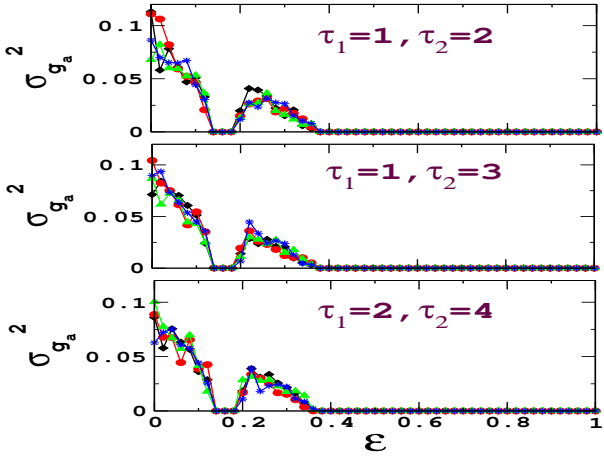}}
\caption{(Color online) ${\sigma_{g_a}}^2$ as a function of $\varepsilon$
for the last generation nodes for 
$N=21$, $K=4$ and for $20$ random initial conditions. The different symbols correspond to the different set of last generation siblings. The local dynamics of the nodes is governed by the logistic map($x_i^{t+1} = 4 x_i^t(1-x_i^t)$).}
\label{fig_Var}
\end{figure}

In this section we discuss lag synchronization of the last generation nodes originated from the same parent in the presence of heterogeneity in delay values.
In order to investigate the lag synchronization we define  
the variance: 
\begin{equation}
{\sigma_{g_a}}^2 = \frac{\langle \sum_{j=a}^N A_{aj}(x_j^{\tau_{ja}}-\bar{x} )^2\rangle_t}{K};   \nonumber
\end{equation}
where $i, j$ are the last generation nodes which
have originated from $a$, $\langle \rangle_t$ denotes average over time and:
\begin{equation}
\bar{x} = \frac{\sum_{j=a}^N A_{aj}x_j^{\tau_{ja}}}{K}. \nonumber
\end{equation}
Thus 
${\sigma_{g_a}}^2 =0   $ for $ x_i^{t+\bigtriangleup\tau} = x_j^{t}$, where $\bigtriangleup\tau  = \tau_1-\tau_2$.
For a network of height $h$ and branch ratio $K$, there will be $K^{h-1}$ set of last generation siblings (represented by g),
 thus $K^{h-1}$ number of variance should 
be calculated, however the behavior of one set of siblings should be same as the other set of siblings. 
Fig.~\ref{fig_Var} manifests variation of $ {\sigma_{g_a}}^2$ vs $\varepsilon$ for the dynamics governed by Eq.~\ref{cml}. 
It presents the lag synchronization among the last generation siblings in both lower
($0.18 \lesssim \varepsilon \lesssim 0.38$) and higher ($\varepsilon\gtrsim 0.38$) coupling range.
 
In order to understand the destruction of exact synchronization and origin of lag synchronization with heterogeneous delay values, 
we study the difference variable between a pair of last generation nodes originating from the same parent
 (let i and j) for the simplest case of $\varepsilon=1$:
\begin{equation}
x_{i}^{t+1} - x_{j}^{t+1} = g(x_{a}^{t-\tau_{ia}}) - g(x_{a}^{t -\tau_{ja}});
\label{diff_var}
\end{equation}
So one can see that for homogeneous delay, the above equation will reduce to :
\begin{equation}
x_{i}^{t+1} - x_{j}^{t+1} = 0;       \nonumber
\end{equation}
Thus, for the homogeneous delay the last generation nodes originating from the same parent
 will always get synchronized, while for the heterogeneous delay at $\varepsilon=1$  a simple calculation gives that the dynamical 
evolution of the $i^{th}$ and $j^{th}$ last generation nodes will be:
\begin{equation}
x_{j}^{t+\bigtriangleup\tau} - x_{i}^t=0; 
\label{eq_diff}
\end{equation}
where $\bigtriangleup \tau = \tau_{ia}-\tau_{ja}$.

\begin{figure}[h]
\includegraphics[width=7cm,height=6cm]{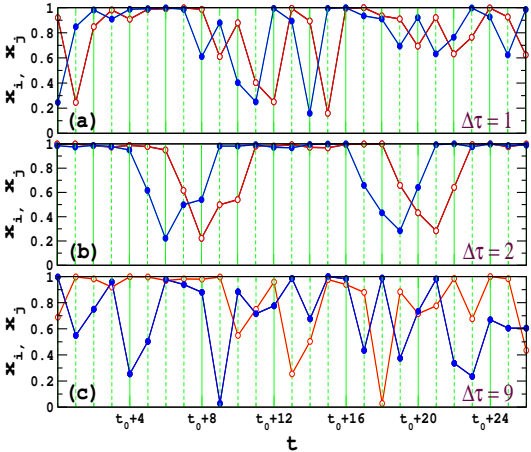}
\caption{(Color online)Time evolution of the boundary nodes originated from the same parent for a
Cayley tree network with $N=21$, $K=4$ and $\varepsilon=1$.
The diagram exhibits that there is time lag synchronization between the
 two nodes i (open circle) and j (closed circle) with time lag being 1, 2 and 9  for
(a), (b) and (c) respectively. 
The local dynamics is governed by the logistic map($x_i^{t+1} = 4 x_i^t(1-x_i^t)$).}
\label{fig_time}\end{figure}

Thus, the introduction of heterogeneity in delay destroys the exact synchronization between a pair of last generation nodes and leads to the 
lag synchronization with time lag being equal to the difference of delay values for the two nodes.
Fig~\ref{fig_time} represents the time evolution of the last generation nodes from the same parent. 

\section{Circle Map}
This section presents results of circle map on Cayley tree. The local dynamics is given by
\begin{equation}
f(x) = x+w+(\frac{\kappa}{2\pi})sin(2\pi x).    \,\,\,\,\,\,\,\,\,   (mod 1)
\label{Circle_map}
\end{equation}
Here we discuss the results with the
parameters of the circle map in the chaotic region ($\omega= 0.44$
and $\kappa = 6$).

\begin{figure}[t]
\includegraphics[width=8cm,height=4cm]{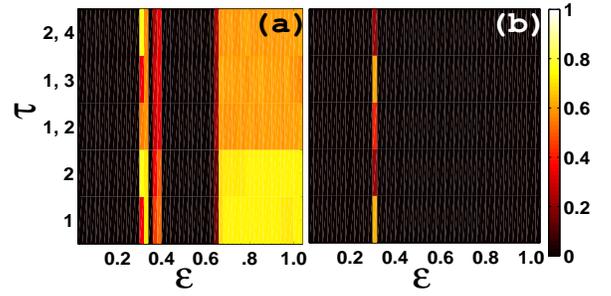}
\caption{(Color online)Phase diagram for the coupled circle maps for Cayley 
tree network of $N=341$ and $K=4$. Shades (colors) stand same as for Fig.(1)}
\label{fig_phase2}
\end{figure}

The circle map at weak coupling represents the change in the mechanism of the cluster formation as observed for the coupled logistic maps 
but for a narrow coupling range ($0.24\lesssim \epsilon \lesssim 0.25$). At strong couplings, the clusters are obtained only through the D mechanism (Fig.~\ref{fig_phase2}), whereas for the logistic map, SO mechanism also played a role in cluster formation (Fig.~\ref{fig_phase}).  The study of the clusters reveal that
the ideal D clusters comprise of only the last generation nodes originating 
from the same parent as demonstrated by the logistic map.

\begin{figure}[h]
\centerline{\includegraphics[width=6cm,height=3cm]{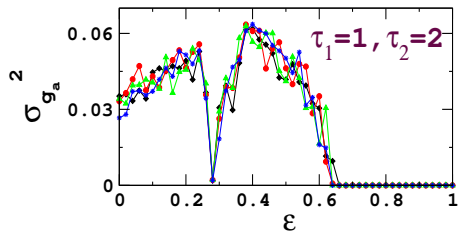}}
\caption{(Color online) ${\sigma_{g_a}}^2$ as a function of coupling strength
for the last generation nodes for circle map.
The figure is plotted for $N=21$, $K=4$ and for 40 random initial conditions. The different symbols correspond to the different set of last generation siblings. The local dynamics of the nodes is governed by Eq.~\ref{Circle_map}.
}
\label{fig_Var2}
\end{figure}

Furthermore to study the lag synchronization of the last generation siblings for diffusively coupled circle maps we plot $ {\sigma_{g_a}}^2$ vs $\varepsilon$ (Fig.~\ref{fig_Var2}). The figure manifests that the lag synchronization of the last generation nodes is observed above a threshold value $\varepsilon>0.62$.

In the case of diffusive coupling, at strong coupling strength, the 
coupling term dominates over the local dynamics. This can be a plausible 
reason behind the lag synchronization of the last generation siblings having a 
common coupling environment.

\section{Results and discussion}
We have studied cluster synchronization on the diffusively coupled Cayley tree networks 
in the presence of heterogeneity in delay values for logistic map and circle map.
We demonstrate that the boundary and the inner nodes in the Cayley tree networks exhibit 
different behavior. The inner nodes get phase synchronized only for the weak coupling, 
while the boundary nodes get synchronized for the weak as well as strong couplings. 
At weak couplings the synchronization of different generations depends on the parity of heterogeneous delay values: (1) The nodes corresponding to alternate
generations get synchronize forming D clusters for even heterogeneous delays, 
(2) the nodes from all the generations or all the nodes in a sub-family
synchronize forming SO clusters
for the odd heterogeneous
delay values and (3) the nodes from the different generations get synchronized 
forming dominant D or mixed clusters for the odd-even heterogeneous delays.
For the first case, there may be synchronization in consecutive generations
as well but in such a manner that D clusters are formed, which is possible
when there is no synchronization between children and parent nodes.

The synchronization between different generations can be analogous to
the inheritance of genetic diseases across generations. 
The SO synchronization can be compared with the disease occurring in all the 
generations of a family or sub-family, while formation of D clusters is similar to  the genetic diseases skipping one or several generations \cite{gen_disease}. 
Although the Cayley tree network model investigated here does not depict
 replica of the genetic disease model, as first of all it represents
single parent tree network, 
and secondly gene network interactions are
far more complex \cite{gene_complex} than the simple model considered here,  the fact
that heterogeneous delay among interacting partners of
infected node (genes) plays a crucial role in occurrence of the synchronization (disease)
in succeeding generation might shed some
light on the understanding of occurrence of genetic diseases \cite{gen_disease2}.
For example, heterogeneous delay
in epistatic gene interactions between the genes inherited from
the diseased ancestor and the genes existing in the individual might be one
of the factors deciding the occurrence of disease in the individual.

At the  intermediate couplings the heterogeneity in delay leads to the synchronization between the  nodes from different generations further leading to an enhancement in the synchronization between the child nodes originating from different parent nodes. This behavior is more enduring in the light that homogeneous delay displays synchronization between the last generation siblings originating from the same parent only.
In addition, the heterogeneity in delay leads to synchronization between the 
parents irrespective of the synchronization among their children nodes, whereas in case of  the homogeneous delays, the parent nodes get synchronized only when children nodes get synchronized \cite{EPJST2013}. 
This indicates that a heterogeneous delay in communication among the members of the family can be more advantageous for the family business \cite{family_business} as it brings harmony between the different generations and disputes among their children do not 
affect the harmony between their parents.
At strong couplings, only the last generation nodes lying on the boundary get synchronized.
It's comparison with homogeneous delay implicates that heterogeneity in delay suppresses the phase synchronization among the siblings of the last generation by distributing them in to different clusters.
Furthermore, the boundary nodes which constitute the maximum population
of Cayley tree demonstrate a completely different behavior than the inner nodes.
The boundary nodes originating from the same parent, receive same 
information as they are connected with their parent node only and thus
lead to the exact synchronization in case of the homogeneous delays. The 
exact synchronization gets destroyed by the heterogeneous delays and  lag synchronization is displayed with the time lag being equal to the difference in the delay value for the corresponding nodes.
The Lyapunov function analysis demonstrates that the destruction of the same coupling environment for the last generation siblings is the cause behind the suppression of the exact synchronization. 
Another work involving delay have shown occurrence of lag synchronization\cite{lag_syn}, where all the nodes of a network show lag synchronization. Our work presents to a new phenomenon by exhibiting that even though the whole system does not display the lag synchronization, the boundary nodes leads to the formation of lag synchronized clusters.
To conclude the heterogeneity
supports cordial behavior among different families as reflected
from the clusters containing nodes originated from different families.
We have demonstrated that heterogeneity in delay leads to the phase synchronized clusters consisting of the nodes from  the different generations and lag synchronized clusters comprising of the nodes from last generation for Cayley tree networks. 
The results presented here have been related with the family business and
have been discussed in the context of genetic diseases, however, the framework presented here needs to incorporate a more realistic
interaction as well as evolution pattern in order to represent these systems \cite{gene_complex,epilepsy,family_business2}. 


\section{Acknowledgments}SJ acknowledges DST and CSIR project grants
SR/FTP/PS-067/2011 and 25(02205)/12/EMR-II for financial support and
R. E. Amritkar for useful suggestions. We thank members of Complex Systems Lab at IITI for providing conducive environment and for useful discussions. AS Acknowledges Camellia Sarkar for useful discussions.

\end{document}